\documentclass{webofc}
\usepackage[varg]{txfonts}
\usepackage[utf8]{inputenc}
\usepackage{hyperref}
\hypersetup{
    colorlinks=true,
    linkcolor=blue,
    filecolor=blue,      
    urlcolor=cyan,
    }
    
\begin{document}

\title{Exploring the Phase Diagram of V-QCD with Neutron Star Merger Simulations}

\author{
\firstname{Tuna} \lastname{Demircik}\inst{1}
\and
\firstname{Christian} \lastname{Ecker}\inst{2}\fnsep\thanks{\email{ecker@itp.uni-frankfurt.de}}
\and
\firstname{Matti} \lastname{J\"arvinen}\inst{3,4}
\and
\firstname{Luciano} \lastname{Rezzolla}\inst{2,5,6}
\and
\firstname{Samuel} \lastname{Tootle}\inst{2}
\and
\firstname{Konrad} \lastname{Topolski}\inst{2}
}

\institute{Department of Theoretical Physics, Wroc\l aw University of Science and Technology, 50-370 Wroc\l aw, Poland
\and
Institut f\"ur Theoretische Physik, Goethe Universit\"at, Max-von-Laue-Str. 1, 60438 Frankfurt am Main, Germany
\and
Asia Pacific Center for Theoretical Physics, Pohang, 37673, Korea
\and
Department of Physics, Pohang University of Science and Technology, Pohang, 37673, Korea
\and
Frankfurt Institute for Advanced Studies, Ruth-Moufang-Str. 1,60438 Frankfurt, Germany
\and
School of Mathematics, Trinity College, Dublin 2, Ireland
}

\abstract{
Determining the phase structure of Quantum Chromodynamics (QCD) and its Equation of State (EOS) at densities and temperatures realized inside neutron stars and their mergers is a long-standing open problem. 
The holographic V-QCD framework provides a model for the EOS of dense and hot QCD, which describes the deconfinement phase transition between a dense baryonic and a quark matter phase. We use this model 
in fully general relativistic hydrodynamic (GRHD) simulations to study the formation of quark matter and the emitted gravitational wave signal of binary systems that are similar to the first ever observed neutron star merger event GW170817.
}

{
\hfill APCTP Pre2022 - 026
}

\maketitle

\section{Introduction}\label{intro}
Multi-messenger observations from binary neutron-star mergers are expected to provide new insights into properties of dense and hot Quantum Chromodynamics (QCD) that are currently inaccessible with first-principle methods in QCD and collider experiments on earth.
A key quantity to establish the connection between the properties of dense QCD matter and neutron stars is the Equation of State (EOS), which in the zero temperature limit determines the pressure as a function of the energy density as it is required to solve for the mass-radius relation of static and cold stars in General Relativity.
Robust predictions for the EOS are only available from chiral effective theory (CET) calculations~\cite{Hebeler:2013nza,Gandolfi2019,Keller:2020qhx,Drischler:2020yad} at densities below and close to the nuclear saturation density $n_s=0.16\,\rm fm^{-3}$ and from perturbative QCD~\cite{Fraga2014,Gorda:2021kme,Gorda:2021znl} at densities that are much larger than those reached in even the most massive neutron stars.
At densities a few times larger than $n_s$, which are realized in the core regions of neutron stars and their mergers, the available options are either model-building or model agnostic parameterizations of the EOS.
The latter approach, when combined with theory input from CET and perturbative QCD together with observational data of neutron star masses and radii~\cite{NANOGrav:2019jur,Fonseca:2021wxt,Riley:2019yda,Miller:2019cac,Nattila:2017wtj,Miller:2021qha,Riley:2021pdl} and bounds on the tidal deformability derived from direct gravitational wave detections of binary neutron star mergers~\citep{Abbott2018a}, has lead to a number of important insights on the cold EOS at neutron star densities~\cite{Annala2019, Annala:2022, Altiparmak:2022} and on the properties of isolated neutron stars (see~\cite{Ecker:2022,Ecker:2022dlg} and references therein).
However, in neutron star mergers, where strong shocks during and after the collision can locally heat matter to temperatures $T>50$~MeV, the zero temperature approximation of the EOS becomes insufficient, which leaves model-building as the only feasible option.

An important open question is whether a phase transition between confined hadrons and deconfined quark matter can happen during and after the collision of two neutron stars.
It is for example expected that imprints of the deconfinement phase transition might be visible in the kHz gravitational-wave signal emitted during a possible post-merger hyper-massive neutron star (HMNS) stage that encodes information about the hot and dense part of the EOS that is inaccessible during the inspiral phase~\cite{Bauswein:2018bma,Weih:2019xvw}.
Addressing the question of the detectability of quark matter in binary neutron-star mergers requires two important ingredients. 
The first one is a model for dense matter that is able to make predictions for the phase structure of QCD at all temperatures and densities realized in neutron-star mergers and that can consistently describe the deconfinement phase transition.
In addition, the EOS derived from such a model needs to be consistent with the aforementioned theoretical constraints obtained from CET and perturbative QCD in their respective regimes of validity, as well as with available observational data of neutron stars and their mergers.
The second crucial ingredient is a numeric framework that is able to use such a EOS model as an input and that can solve
% for the complicated general relativistic hydrodynamic time evolution of 
complicated GRHD equations of a neutron star merger as well as extract the gravitational waves produced in such events.
In the following, we shall use EOSs that were constructed with the recently developed 
finite-temperature 
framework based on the holographic V-QCD model~\cite{Demircik:2021zll} 
together with a set of state-of-the-art general relativistic hydrodynamics codes~\cite{Most2019b,Papenfort2021b} to simulate the formation of quark matter in the HMNS phase of binary neutron star mergers that resemble the first-ever detected event GW170817.

This proceedings contribution summarises material published in~\cite{Demircik:2021zll,Tootle:2022pvd} and was presented at \textit{The XVth Quark confinement and the Hadron spectrum conference}.
%Section~\ref{sec-1} gives a brief introduction to the V-QCD framework.
%Section~\ref{sec-2} provides details on the numerical implementation of the binary neutron star merger simulations performed.
%In Section~\ref{sec-3} we present the key results of these simulations. 
%Finally, in Section~\ref{sec-3} we summarise.
%Further details on the EOS construction and the numeric merger simulations can be found in~\cite{Demircik:2021zll} and \cite{Tootle:2022pvd}, respectively.

\section{The V-QCD Equation of State Framework} 
\label{sec-1}
\begin{figure*}[htb]
\center
    \includegraphics[height=0.34\textwidth]{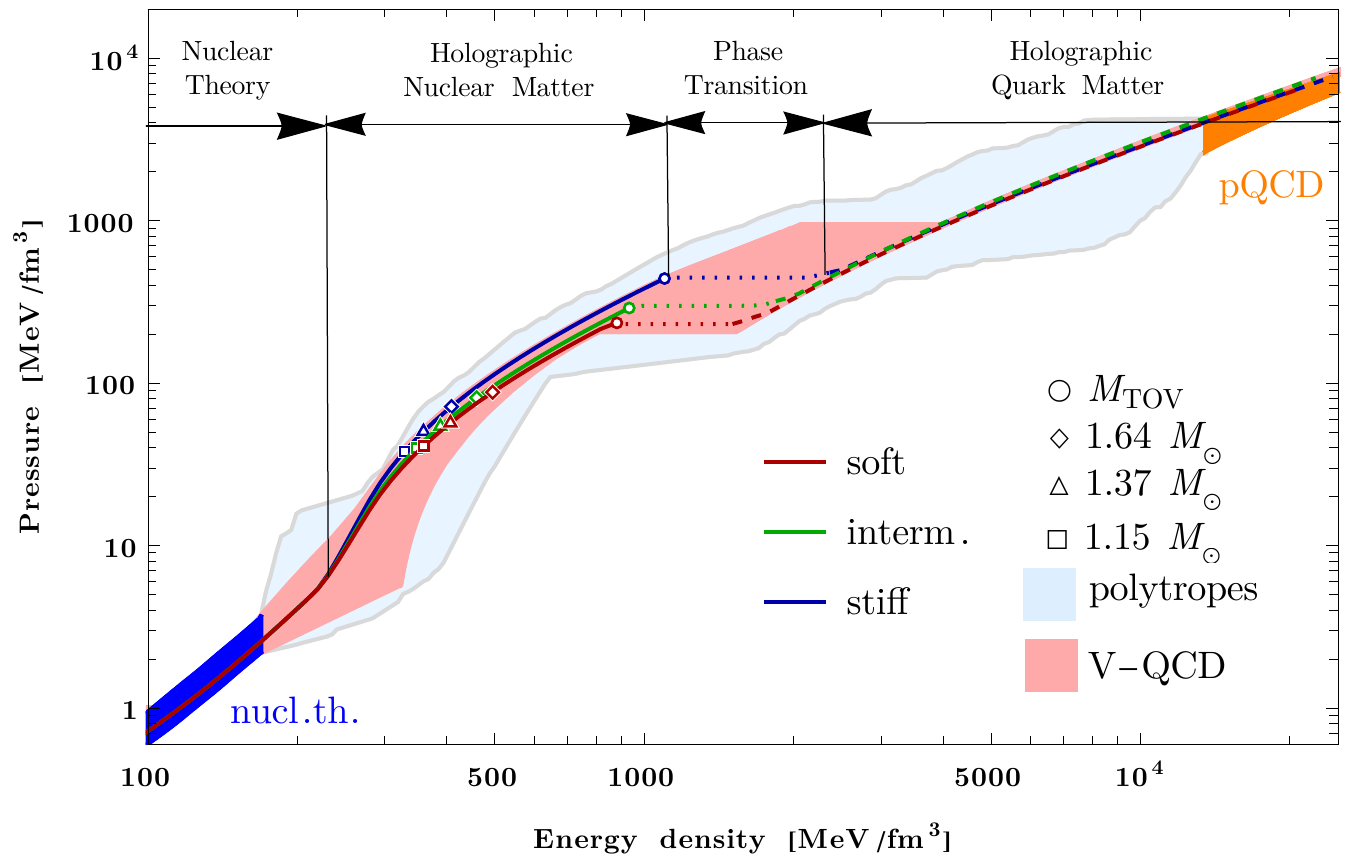}\quad
    \includegraphics[height=0.34\textwidth]{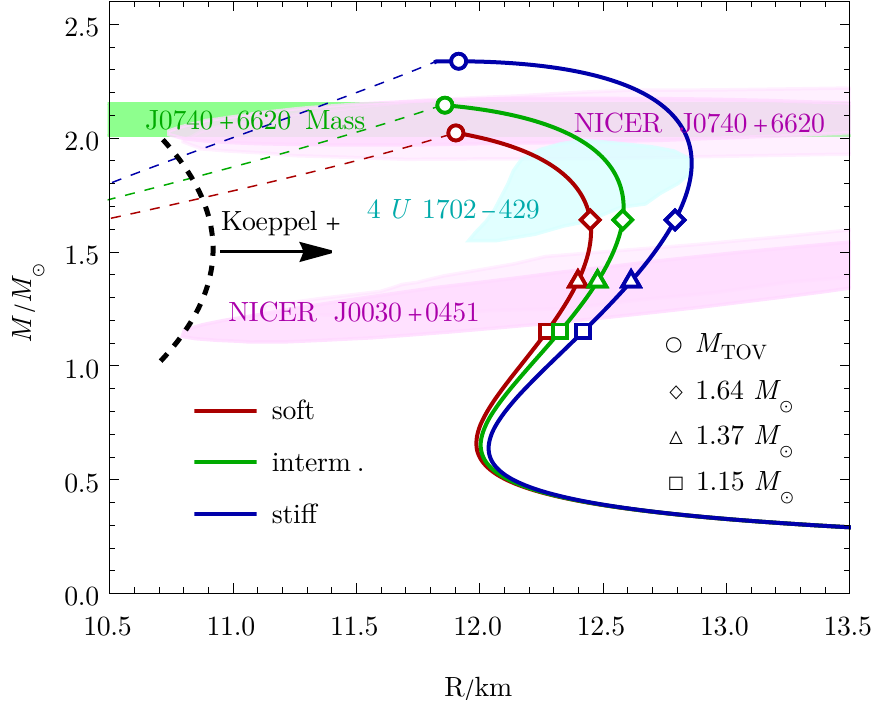}
    \caption{\small Left: Cold, beta equilibrium slices of the V-QCD
      EOS. Shown are the soft (red), intermediate (green) and stiff (blue)
      versions of the model. Right: Corresponding mass-radius relations where
      dashed lines mark unstable quark matter branches. 
      See~\cite{Tootle:2022pvd} for details.
     % Dark blue and orange bands mark the
     % uncertainties in nuclear theory \cite{Hebeler:2013nza} and
     % perturbative QCD \cite{Fraga2014} calculations, respectively.
     % The light blue region marks the theoretically allowed region
     % spanned by polytropic EOSs, whereas the pink region marks the
     % parametric freedom of V-QCD after imposing constraints from NS
     % observations. Right: Corresponding mass-radius relations where
     % dashed lines mark unstable quark matter branches. In addition we
     % show various error bands of direct mass~\cite{NANOGrav:2019jur,
     %   Fonseca:2021wxt} and radius~\cite{Riley:2019yda, Miller:2019cac,
     %   Nattila:2017wtj, Miller:2021qha, Riley:2021pdl} measurements of
     % heavy pulsars; shown with a black dashed line is a theoretical
     % lower bound for the radii as computed from considerations on the
     % threshold mass by Koeppel et al.~\cite{Koeppel2019}.
     }
    \label{fig:MR}
\end{figure*}
In this section we briefly summarise the V-QCD framework for the hybrid EOS model used in our merger simulations (for details see~\cite{Jarvinen:2021jbd,Demircik:2021zll}).
The important novel component of this construction is the use of a gauge/gravity duality calculation for the dense nucleonic phase and its transition to deconfined quark matter.
The gauge/gravity duality~\cite{Maldacena:1997re,Witten:1998qj} allows one to translate intractable problems in quantum field theory at strong coupling and finite density to tractable problems in classical five-dimensional gravity and has been recently applied by various groups to neutron-star physics~\cite{Hoyos:2016zke,Ghoroku:2019trx, BitaghsirFadafan:2019ofb,Ecker:2019xrw, Demircik:2020jkc,Kovensky:2020xif,Pinkanjanarod:2020mgi,Mamani:2020pks,Ghoroku:2021fos,Jarvinen:2021jbd, Hoyos:2021uff, Kovensky:2021kzl, Bartolini:2022rkl}.

V-QCD is a holographic bottom-up model with a number of free parameters that are tuned to reproduce characteristic features of QCD, such as confinement and chiral symmetry breaking and are further constrained by demanding consistency with lattice QCD data and astrophysical measurements~\cite{Jarvinen:2021jbd}.
The V-QCD action consists of a gluon part that is given by improved holographic QCD~\cite{Gursoy:2007cb,Gursoy:2007er}, a quark (flavour) part implemented with a tachyonic Dirac-Born-Infeld action~\cite{Bigazzi:2005md,Casero:2007ae} and a component for homogeneous nuclear matter derived 
in~\cite{Ishii:2019gta}. The 
so-called Veneziano limit is assumed in which both the number of colours $N_c$ and flavours $N_f$ is large but their ratio is $\mathcal{O}(1)$~\cite{Jarvinen:2011qe} in order to maintain the backreaction between quarks and gluons in the holographic large $N_c$ limit.
This calculation is then extended to densities around and below $n_s$, where the homogeneous approximation for V-QCD nuclear matter becomes unreliable, with a combination of the traditional nuclear theory APR~\cite{Akmal:1998cf} and HS(DD2)~\cite{Hempel2010,Typel:2009sy} EOSs including their temperature dependence up to $T\approx 160~$MeV, as well as with bosonic contributions given by a black body photon gas and all mesons of mass $\leq 1$~GeV from the particle data group listings~\cite{ParticleDataGroup:2020ssz}. 
The latter contributions are important to model the QCD critical end point of the deconfinement transition. 
The finite temperature extension of the cold V-QCD baryon sector is realised by matching it with a van der Waals hadron gas model~\cite{Rischke:1991ke}, whose details can be found in~\cite{Demircik:2021zll}.
Finally, the quark fraction and the phase boundaries of the mixed phase at the first order deconfinement transition is implemented with a Maxwell construction.
In this way the combined model is able to provide predictions for the finite temperature EOS on the part of the QCD phase diagram that is relevant to binary neutron star mergers, core-collapse supernovae and heavy-ion collision experiments. 

We employ three variants of the EOS, which were introduced in~\cite{Demircik:2021zll} in order to represent the parameter dependence of the model. 
These are called the soft, intermediate, and stiff variants; and reflect three different choices of the parameters in the action of V-QCD~\cite{Jokela:2018ers}.
In Fig.~\ref{fig:MR} (left) we show cold beta-equilibrium slices of these three EOSs with uncertainty bands from nuclear theory~\cite{Hebeler:2013nza} (blue) and perturbative QCD~\cite{Fraga2014} (orange). 
Red, green and blue lines are the soft, intermediate and stiff versions of V-QCD where the dotted part of these curves mark the first order phase transition between the baryonic (solid) the quark matter (dashed) phase.
In addition we show markers for the central densities reached in various isolated non-rotating stars that we use to initialise the binary systems in our simulations. 
Light blue and pink regions mark the residual freedom of model agnostic parametrizations of the EOS~\cite{Annala2019,Annala:2022} and of the V-QCD model~\cite{Jokela:2020piw}, respectively, after imposing constraints from neutron star observations. 
In Fig.~\ref{fig:MR} (right) we show the corresponding mass-radius relations of non-rotating stars together with various error bands of the direct mass measurement (green area) of the heavy pulsar PSR~J0740+6620~\cite{NANOGrav:2019jur,Fonseca:2021wxt} ($M=2.08 \pm 0.07 M_\odot$) and direct radius measurements of PSR~J0740+6620~\cite{Miller:2021qha,Riley:2021pdl} and PSR~J0030+0451~\cite{Riley:2019yda,Miller:2019cac} obtained by the NICER experiment (pink ellipses) as well as from the measurement of the X-ray binary 4U~1702-429~\cite{Nattila:2017wtj} (cyan area; see~\cite{Jokela:2021vwy} for a more detailed analysis of the impact of the NICER results).
Finally, shown with a black dashed line is a theoretical lower bound for the radii as computed from considerations on the threshold mass by Koeppel et al.~\cite{Koeppel2019}.
We note that we have checked that our EOSs respect the upper bound on the binary tidal deformability $\tilde \Lambda<720$ (low-spin priors) obtained from GW171817 and are well within the bounds recently proposed on general parametrization of the sound speed in neutron stars \cite{Altiparmak:2022}.

\section{Numerical Methods}
\label{sec-2}
A crucial component in the numerical simulation of binary neutron star mergers
is accurate initial data for the subsequent time evolution.
For this we use the recently developed Frankfurt University/Kadath
(FUKA)~\citep{Papenfort2021b} code suite, based on 
an extended version of the KADATH spectral solver
library~\citep{Grandclement09}. FUKA is able to generate realistic compact binary systems even with large mass asymmetry and sizeable individual spins.
For the binary evolution we make use of the Einstein Toolkit~\cite{EinsteinToolkit_etal:2020_11} infrastructure that
includes the fixed-mesh box-in-box refinement framework Carpet
\cite{Schnetter-etal-03b}, which keeps track of the location 
of the individual stars and the merger remnant in order to provide a fine 
computational grid where this is needed.
For example, in the simulations discussed below, six refinement levels
with finest grid-spacing of $221\,{\rm m}$ were used.
To solve the fully general relativistic hydrodynamic equations in 3+1 spacetime 
dimensions, we use a number of code components developed by the 
Frankfurt group and that are implemented in the Einstein Toolkit.
For the spacetime evolution we use Antelope~\citep{Most2019b},
which solves a constraint damping formulation of the Z4 system~\citep{Bernuzzi:2009ex,Alic:2011a}.
Furthermore, to evolve the hydrodynamic part we use the
Frankfurt/Illinois (FIL) general-relativistic
magnetohydrodynamic code~\citep{Most2019b}, which is based on the
IllinoisGRMHD code \citep{Etienne2015} and implements
fourth order conservative finite-differencing methods
\citep{DelZanna2007}, enabling a precise hydrodynamic evolution even at
relatively low resolution.
\begin{figure*}[htb]
\centering
    \includegraphics[width=0.85\textwidth]{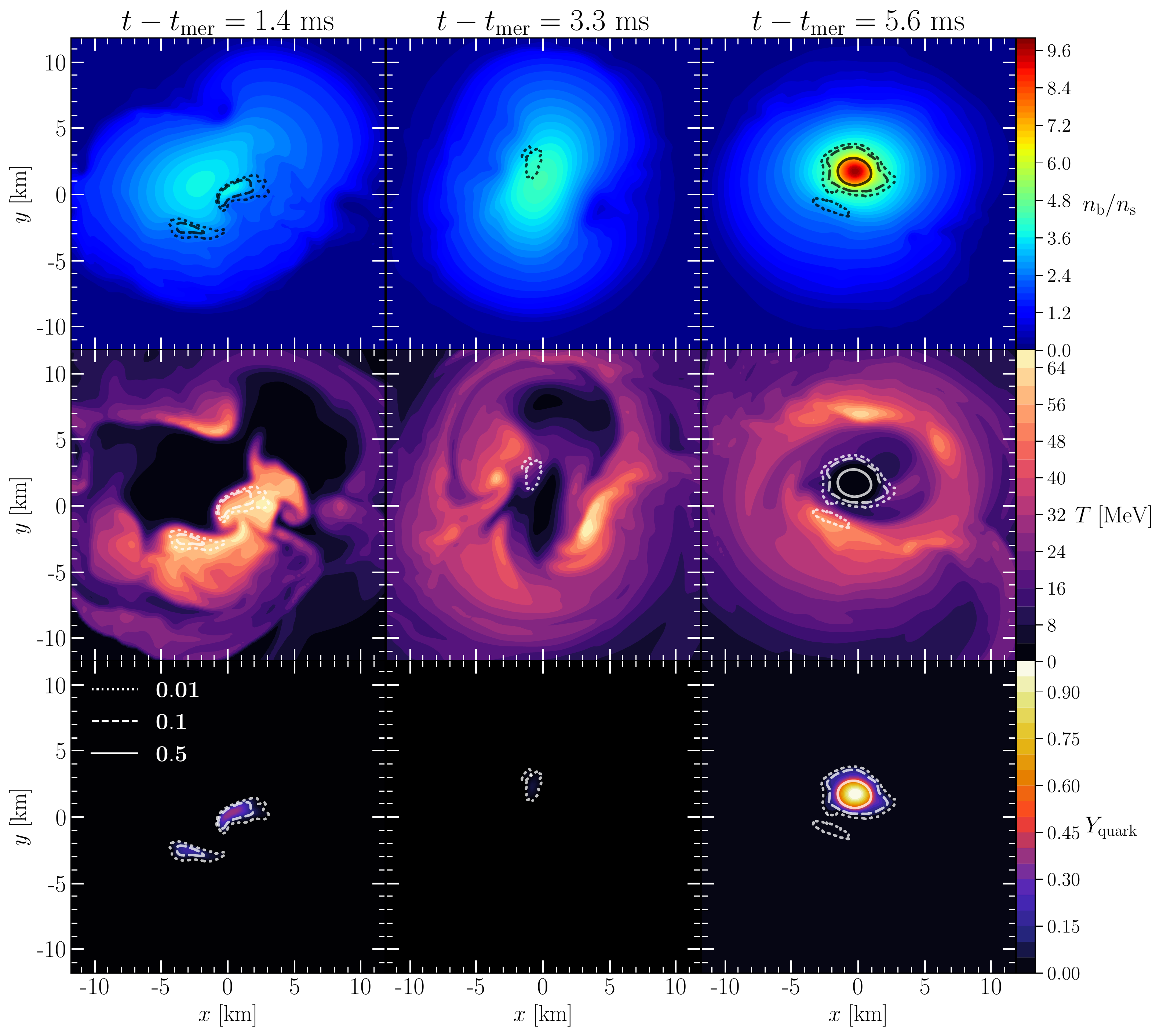}
     \caption{ From top to bottom rows we show snapshots of the
       baryon-number density, temperature, and quark fraction in the
       orbital plane at three different times $t-t_{\rm
         mer}=1.4,\,3.3,\,5.6\,\,{\rm ms}$ (from left to right) that are
       representative for hot, warm and cold quark stages of a $q=0.7$ simulation with soft V-QCD EOS. In addition, we mark contours for the quark fraction
       $Y_{\rm quark}=0.01,\,0.1,\,0.5$ by dotted, dashed and solid
       lines, respectively. See~\cite{Tootle:2022pvd} for details.}
    \label{fig:soft_q0.7_hr_3x3_plot2D}
\end{figure*}

\section{Results}
\label{sec-3}
\begin{figure*}[htb]
    \center
    \includegraphics[width=0.74\textwidth, keepaspectratio]{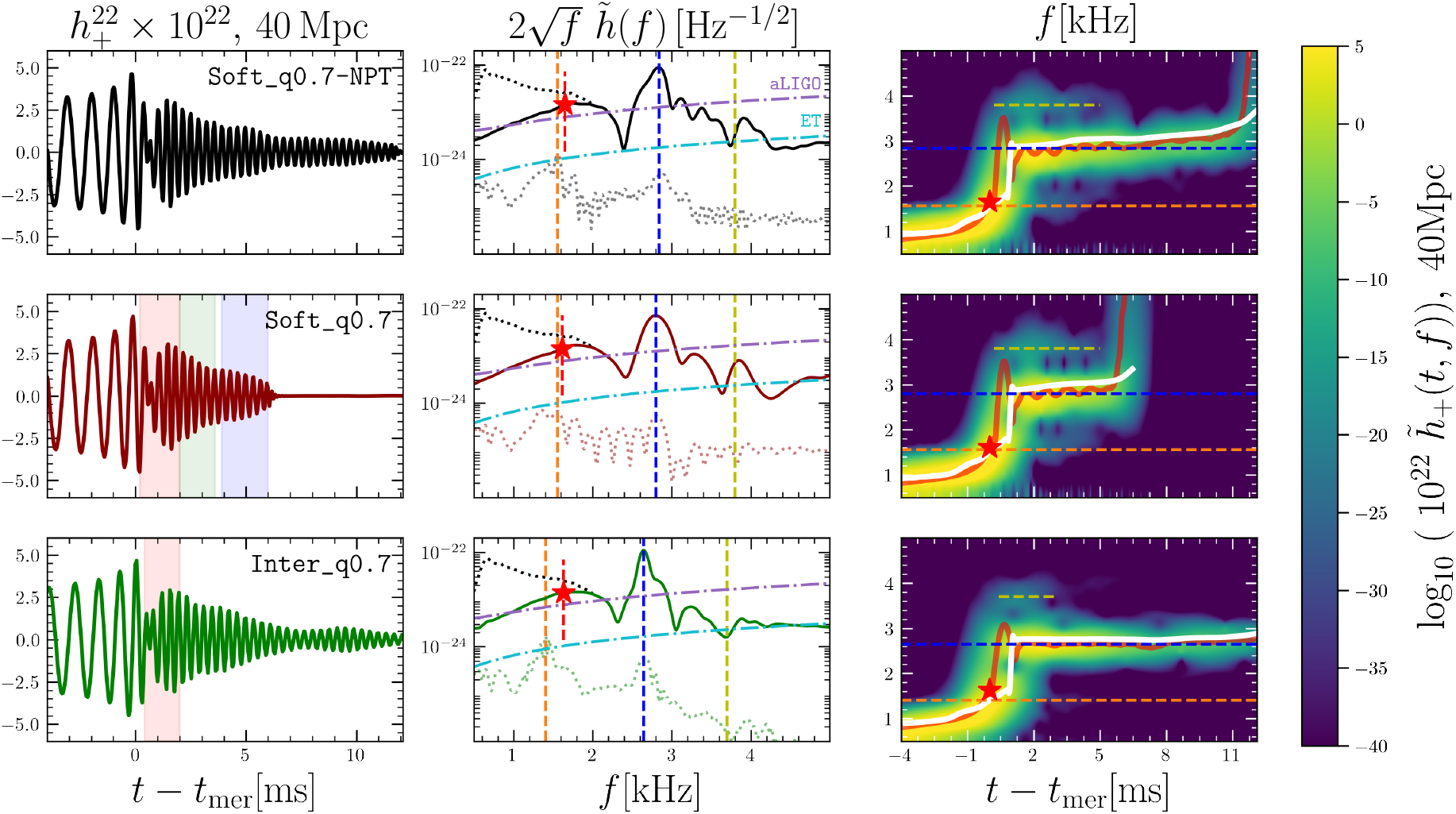}
     \caption{Shown above is the $h^{22}_+$ of the strain (left), the 
       power-spectral density of the effective strain (middle), and the spectrogram (right) of
       the $h^{22}_+$ of the soft model with (top) and without quark matter (center) as well as the intermediate model (bottom) for $q = 0.7$ and $\mathcal{M}_{\rm chirp} = 1.188$.
       See~\cite{Tootle:2022pvd} for details.
     % Dashed blue and orange lines correspond to the $f^{2,2}$ and $f^{2,1}$ peaks respectively, the dashed yellow lines corresponds to the $f_3$ peak as measured in the spectrogram, and the star denotes the peak merger frequency, $f_{\rm mer}$
%       In the right two panels the dashed blue and orange lines
%       correspond to the $f^{2,2}$ and $f^{2,1}$ peaks respectively, the
%      dashed yellow lines corresponds to the $f_3$ peak as measured in
%       the spectrogram, and the star denotes the peak merger frequency, $f_{\rm
%         mer}$. In the right figure, the white line traces
%       the maximum in the spectrogram. Finally, the sensitivity curves in
%       the middle plot are related the current sensitivity of advanced
%       LIGO and the Einstein Telescope respectively
%       \cite{adLIGO2018,Punturo:2010zza}.
    }
    \label{fig:gwtable}
\end{figure*}

We have performed a set of fully general relativistic hydrodynamics simulations of binary neutron star systems with the measured chirp mass $\mathcal{M}_c=\frac{(M_1\,M_2)^{3/5}}{(M_1+M_2)^{1/5}}=1.186\,M_\odot$ of GW170817 and two different ratios of the individual neutron star masses $q=\frac{M_2}{M_1}=0.7,1$.
We simulate these binaries using the aforementioned variants (soft, intermediate, stiff) of the V-QCD model.
In Fig.~\ref{fig:gwtable} we show snapshots of various quantities in the orbital plane of the HMNS at three different times ($1.4-5.6$~ms) after the merger.
The case depicted is the result of an unequal mass binary with $q=0.7$ and soft V-QCD EOS. 
The top row shows the baryon number density $n_b$ in units of $n_s$.
Regions where quark matter is formed are enclosed by dotted, dashed and solid black lines.
The initially highly deformed HMNS becomes more circular and compact as time proceeds until a dense quark matter core forms, which then ultimately leads to a phase transition triggered collapse of the merger remnant into a black hole (not shown).
The second row shows the corresponding temperature distribution.
Here one of our main findings becomes visible.
Right after the collision (first column), strong shocks locally lead to temperatures beyond $50$~MeV inside the HMNS.
This is where hot quark matter is formed in the hottest but not in the densest regions of the HMNS.
The second row shows the situation a few milliseconds later, where a small amount of warm quark matter is formed at intermediate densities and temperature.
In the third row the situation right before the black hole collapse is shown.
Here a massive core of cold pure quark matter is formed in the central region of the HMNS where the density is high, but the temperature is low.
The decrease in temperature of the quark matter core can be explained by the sizeable latent heat that is necessary to complete the transition from baryonic to quark matter.  
Finally, in the third row we report the fraction of quark matter $Y_{\rm quark}$ produced in the early hot, intermediate warm and late cold quark stages of the HMNS.

In Fig.~\ref{fig:gwtable} we show for three representative simulations of $q=0.7$ binaries the waveform and their frequency spectra that are extrapolated to the $40$~Mpc luminosity distance of the GW170817 event.
First, second and third column are the plus-polarization of the dominant $\ell=m=2$ mode of the gravitational wave strain, the power spectral density together with the sensitivities of the advanced LIGO and the Einstein Telescope detectors, and the gravitational wave spectrogram, respectively, where coloured dashed lines indicate various characteristic post-merger peak-frequencies.
The first line shows the results for the soft V-QCD EOS where the quark phase has been excluded, the second line shows the same configuration where the quark phase is included, and the third line is the corresponding result for the intermediate V-QCD EOS.
Comparing these results shows that the dominant effect of quark matter is the early phase transition triggered black hole collapse that leads to an early termination of the wave form (first versus second line).
The post-merger frequencies in all three cases turn out to be almost identical, showing that the three quark matter stages in our simulations have negligible impact on the details of the waveform.
\section{Summary}
\label{sec-4}
We carried out fully general relativistic hydrodynamic simulations of neutron star binary systems that are similar to GW170817. These simulations employ different variants of holographic V-QCD EOSs allowed us to identify three different post-merger stages in which hot, warm and cold quark matter is formed.
The dominant impact of quark matter on the HMNS evolution is a phase transition triggered collapse to a black hole that results in an early truncation of the gravitational wave signal. The impact of quark matter on the post-merger kHz gravitational wave spectrum turns out to be small.
Future work focuses on the impact of quark matter on the threshold mass of promptly collapsing heavy binaries and the long-time stability of lighter binary systems. 
The EOS files used in the simulations are publicly available on the CompOSE database~\cite{Typel:2013rza,Typel:2022lcx}
%~\url{https://compose.obspm.fr/}
under these links \href{https://compose.obspm.fr/eos/290}{DEJ(DD2-VQCD) soft}, \href{https://compose.obspm.fr/eos/289}{DEJ(DD2-VQCD) intermediate}, \href{https://compose.obspm.fr/eos/291}{DEJ(DD2-VQCD) stiff}. 

\section*{Acknowlegments}
CE acknowledges support by the Deutsche Forschungsgemeinschaft (DFG,
German Research Foundation) through the CRC-TR 211 'Strong-interaction
matter under extreme conditions'-- project number 315477589 -- TRR 211.
ST, KT, and LR acknowledge the support by the State of Hesse within the
Research Cluster ELEMENTS (Project ID 500/10.006). TD and MJ have been
supported by an appointment to the JRG Program at the APCTP through the
Science and Technology Promotion Fund and Lottery Fund of the Korean
Government. TD and MJ have also been supported by the Korean Local
Governments -- Gyeong\-sang\-buk-do Province and Pohang City -- and by
the National Research Foundation of Korea (NRF) funded by the Korean
government (MSIT) (grant number 2021R1A2C1010834). T.D. also acknowledges the support of the Narodowe Centrum Nauki (NCN) Sonata Bis Grant No. 2019/34/E/ST3/00405. LR acknowledges
funding by the ERC Advanced Grant ``JETSET: Launching, propagation and
emission of relativistic jets from binary mergers and across mass
scales'' (Grant No. 884631). The simulations were performed on HPE Apollo
HAWK at the High Performance Computing Center Stuttgart (HLRS) under the
grant BNSMIC. For our visualisations we made an extensive use of the
Kuibit library \cite{kuibit21}.

\bibliography{main.bib}

\end{document}